\documentclass[twoside]{article}
\usepackage{epsfig}

\begin{document}
31 August 2016 \\

\textbf{{\Large FO Aqr time keeping}} \\

Michel Bonnardeau

\textit{MBCAA Observatory, Le Pavillon, 38930 Lalley, France, email: arzelier1@free.fr} \\

\textit{\textbf{Abstract: Twelve seasons, from 2004 to 2015, of photometric monitoring of the intermediate polar FO Aqr are presented and are compared with previous observations. The ambiguities in the cycle counting can be lifted and a new O-C diagram, spanning 34 yr, is presented, along with new ephemerides.}} \\ \\
\\
\textbf{}
\\
\textbf{Introduction} \\

FO Aquarii (RA=22h 17min 55.39s DEC=-08$^{\circ}$21' 03.8", J2000) is an intermediate polar, that is a subclass of cataclysmic systems in which the white dwarf is magnetized enough to module the accretion. Furthermore, the period of rotation (or spin) of the white dwarf is shorter than the orbital period. FO Aqr is one of the brightest of its kind. 

The orbital period of FO Aqr is $P_{orb}=4.85$ hr, the rotation (or spin) period of the white dwarf is $P_{rot}=1254$ s (Patterson et al, 1998, hereafter P98). They  are visible as modulations in optical photometry, the rotation modulation being fairly strong with an amplitude of 0.2 mag. Some faint and random sideband modulations may show up; however, following e.g. P98 and Williams, 2003 (hereafter W03), they will be neglected in the analysis that will follow.

Twelve seasons of photometric monitoring, from 2004 to 2015, are to be presented and compared with previous observations. The analysis is similar to the one in Bonnardeau, 2015 for AO Psc.
\\ \\
\textbf{}
\\
\textbf{Observations} \\

The observations were carried out with a 203 mm f/6.3 Schmidt-Cassegrain telescope, a Clear filter and a SBIG ST7E camera (KAF401E CCD). The exposures were 60 s long. For the differential photometry, the comparison star is GSC 5803-398. A total of 5328 useful images were obtained over 49 nights. Figure 1 shows an example of a light curve.
\\
\\ \\ \\ \\ \\ \\
\textbf{Analysis of the modulations} 
 
The magnitudes as a function of time t are fitted by the following H(t) function: 
\\
$H(t) = A_{0} + H_{rot}(t) + H_{orb}(t) $
\\
where $A_{0}$ is a constant, $H_{rot}(t)$ is the rotation modulation: 
\\
$H_{rot}(t) = A_{rot}[cos(\pi(t - t_{rot})/P_{rot}]^2$ 
\\
and $H_{orb}(t)$ is the orbital modulation: 
\\
$H_{orb}(t) = A_{orb}[1+cos(2\pi(t-t_{orb})/P_{orb})]$ 

The H(t) function is fitted to the observations using a Monte Carlo method to test the parameters relative to the timing, and, for each trial, the amplitudes are determined by a least-squares method. The fits are weighted with the uncertainties on the observations. This is done season by season and the results are listed in Table 1. 
\vskip 1.0cm
\epsfysize=6.0cm
\centerline{\epsffile{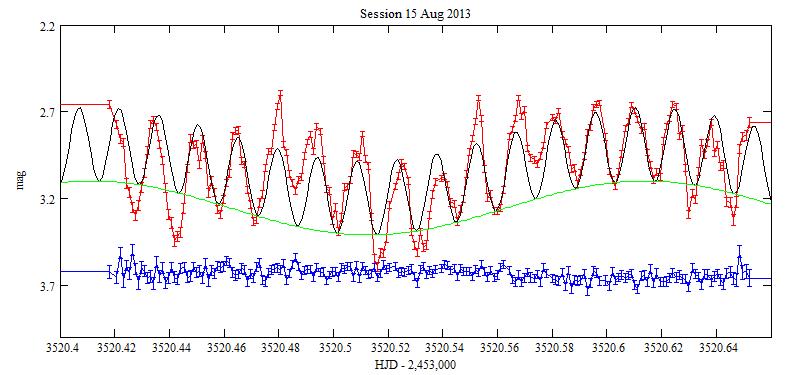}}
\textit{Figure 1: Upper light curve: FO Aqr, lower one: the check star UCAC4 409-138161. The error bars are the quadratic sum of the 1-sigma statistical uncertainties on the comparison star and, respectively, on  the variable star or the check star. Dark line: $H(t)$ function, green line:  $H_{orb}(t)+A_{0}$ function.}

\begin{center}
\begin{tabular}{|l|r|r|r|r|r|r|r|r|r|r|}
\hline {\small Season}  & $t_{orb}$ & $\Delta N_{orb}$ & $t_{rot}$ & $\Delta N_{rot}$ & $A_{0}$
 & $A_{orb}$ & $A_{rot}$\\
  \hline {\small 2004} & {\small 249.55091 } & {\small (a)} & {\small $245.48796$} & {\small (b)} & {\small $2.828$} & {\small $0.103$} & {\small $-0.194$} \\
  & {\small $\pm 0.00069$} & & {\small $\pm 0.00007$} & & {\small $\pm 0.029$} & {\small $\pm 0.001$} & {\small $\pm 0.004$} \\
      \hline {\small 2005} & {\small $695.29533 $} & {\small 2206} & {\small $695.37821$} & {\small 30988} & {\small $2.864 $} & {\small $0.126$} & {\small $-0.213$} \\
        & {\small $\pm 0.00042$} & & {\small $\pm 0.00008$} & & {\small $\pm 0.039$} & {\small $\pm 0.008$} & {\small $\pm 0.008$} \\
   \hline {\small 2006} & {\small $916.5548$} & {\small 1095} & {\small $999.44714$} & {\small 20944} & {\small $2.932$} & {\small $0.058$} & {\small $-0.237$} \\
       & {\small $\pm 0.0012$} & & {\small $\pm 0.00006$} & & {\small $\pm 0.044$} & {\small $\pm 0.001$} & {\small $\pm 0.009$} \\
                  \hline {\small 2007} & {\small $1294.6062 $} & {\small 1871} & {\small $1350.43570$} & {\small 24176} & {\small $2.887$} & {\small $0.059$} & {\small $-0.242$} \\
      & {\small $\pm 0.0013$} & & {\small $\pm 0.00009$} & & {\small $\pm 0.032$} & {\small $\pm 0.004$} & {\small $\pm 0.006$} \\
         \hline {\small 2008} & {\small $1682.56750$} & {\small 1920} & {\small $1682.55057$} & {\small 22876} & {\small $2.763$} & {\small $0.137$} & {\small $-0.234$} \\    & {\small $\pm 0.00061$} & & {\small $\pm 0.00005$} & & {\small $\pm 0.016$} & {\small $\pm 0.001$} & {\small $\pm 0.002$} \\
\hline {\small 2009} & {\small $2058.60433 $} & {\small 1861} & {\small $2058.59430$} & {\small 25902} & {\small $2.864$} & {\small $0.148$} & {\small $-0.194$} \\    & {\small $\pm 0.00056$} & & {\small $\pm 0.00010$} & & {\small $\pm 0.022$} & {\small $\pm 0.003$} & {\small $\pm 0.003$} \\
\hline {\small 2010} & {\small $2478.49376$} & {\small 2078} & {\small $2416.51837$} & {\small 24654} & {\small $2.749$} & {\small $0.147$} & {\small $-0.220$} \\    & {\small $\pm 0.00030$} & & {\small $\pm 0.00010$} & & {\small $\pm 0.013$} & {\small $\pm 0.006$} & {\small $\pm 0.003$} \\
\hline {\small 2011} & {\small $2745.5946$} & {\small 1322} & {\small $2794.59157 $} & {\small 26042} & {\small $2.835$} & {\small $0.083$} & {\small $-0.209$} \\    & {\small $\pm 0.0035$} & & {\small $\pm 0.00015$} & & {\small $\pm 0.020$} & {\small $\pm 0.001$} & {\small $\pm 0.003$} \\
\hline {\small 2012} & {\small $3125.4822$} & {\small 1880} & {\small $3167.42265$} & {\small 25681} & {\small $2.887$} & {\small $0.102$} & {\small $-0.261$} \\    & {\small $\pm 0.0014$} & & {\small $\pm 0.00005$} & & {\small $\pm 0.019$} & {\small $\pm 0.001$} & {\small $\pm 0.003$} \\
\hline {\small 2013} & {\small $3520.5132$} & {\small 1955} & {\small $3520.43653$} & {\small 24316} & {\small $2.900$} & {\small $0.121$} & {\small $-0.331$} \\    & {\small $\pm 0.0014$} & & {\small $\pm 0.00010$} & & {\small $\pm 0.023$} & {\small $\pm 0.002$} & {\small $\pm 0.005$} \\
\hline {\small 2014} & {\small $3830.4561$} & {\small 1534} & {\small $3887.58892$} & {\small 25290} & {\small $2.957$} & {\small $0.021$} & {\small $-0.288$} \\    & {\small $\pm 0.0069$} & & {\small $\pm 0.00005$} & & {\small $\pm 0.020$} & {\small $\pm 0.001$} & {\small $\pm 0.004$} \\
\hline {\small 2015} & {\small $4214.5996$} & {\small 1901} & {\small $4327.40413$} & {\small 30295} & {\small $2.903$} & {\small $0.100$} & {\small $-0.359$} \\    & {\small $\pm 0.0015$} & & {\small $\pm 0.00005$} & & {\small $\pm 0.020$} & {\small $\pm 0.001$} & {\small $\pm 0.004$} \\
\hline
\end{tabular}
\end{center}
\textit{Table 1: $t_{xxx}$ in BJD - 2,453,000 \\
$\Delta N_{xxx}$ number of cycles from the previous season \\
(a) 12,478 cycles from the last observation of P98, 41,902 cycles from the origin of P98.\\
(b) 582,866 cycles from the origin of W03, see text. }
\\ 
\textbf{}
\\
\textbf{Analysis of the orbital minima} \\
 
There are 70 orbital minima available: 54 from P98, 3 from Kruszewski \& Semeniuk, 1998, 1 from Kennedy et al, 2016, and 12 from this work. They span 34 yr and 61,525 orbits. 

These minima are converted in BJD using the on-line tool of the University of Ohio. They are fitted with the linear ephemeris $t_{orb}(e)=T_{orb}+P_{orb}.e$ using a Monte Carlo method. The results are:
\\
$T_{orb} = 2444782.870469 \pm 0.000056$ BJD
\\
$P_{orb} = 0.2020594932 \pm 0.000000002$5 day

This is compatible with the ephemeris of P98, with an improved precision. The resulting O-C diagram shows only noise.
\\ \\
\textbf{}
\\
\textbf{Analysis of the rotation maxima} \\ 

The last observations before my measurements are the ones from W03 in 1992, 1997, 1998, 2001 and 2002. The cycle count for these measurements can be computed from the ephemeris of W03: 281,276 for 1992, 405,380 for 1997, 433,342 for 1998. For 2001 the cycle count is ambiguous, it is either 510,202 or 510,203. The number of cycles between 2001 and 2002 is 21,089.

The number of cycles between the 2002 observation of W03 and my first measurement, in 2004, is also ambiguous: it is either 51,574 or 51,573.

I fit together the W03 observations and mine with a quadratic ephemeris $t_{rot}(e)=T_{rot}+P_{rot}.e+b_{rot}.e^2$. Because of the ambiguities in the cycle counting, there are 4 different ways to do the fit. The residuals of the 4 fits are shown in the O-C diagram of Figure 2.
\\ \\
\epsfysize=7.0cm
\centerline{\epsffile{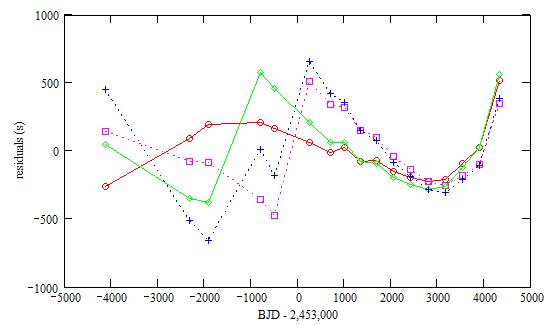}}
\textit{Figure 2: Red circles: the residuals for 2001 at cycle number 510,203 and 2004 at 51,574 cycles from 2002, \\
 Magenta squares with dot line: same for 2001 and 2004 at 51,573 cycles from 2002, \\
  Green diamonds: 2001 at cycle number 510,202 and 2004 at 51,574 cycles from 2002, \\ 
  Blue crosses with dot line: same for 2001 and 2004 at 51,573 cycles from 2002. }
  \\ 
  
The fit that gives the least residuals and the smoothest ones is for the 2001 measurement of W03 at cycle number 510,203 and for 51,574 cycles between the 2002 measurement and my 2004 measurement. \\

Actually, W03 considered 3 different values for the cycle count for the 2001 measurement (and not 2). So I also considered the 2001 measurement with the cycle count numbers 510,201 and 510,204. The 4 extra fits give residuals that are larger than the ones shown in the Figure 2. 

So, by interpolation, the ambiguities of the cycle count are lifted.
\\ 

There are 140 rotation maxima available: \\ 
114 from P98 and 7 from Kruszewski \& Semeniuk (1998) whose cycle numbers can be calculated from the ephemeris of W03; \\
1 from Andronov et al (2005) (there are a V measurement and a R measurement and a combination of both; I used this last one), at -1259 cycles from my 2004 observation and 581,607 cycles from the origin of W03; \\
1 from Kennedy et al (2016) at 8,601 cycles from my 2014 observations and 842,336 cycles from the origin of W03; \\
5 from W03 and 12 from this work, whose cycle numbers have been determined above. \\ These data span 34 years and 864,030 rotations.

The 140 times of maxima are converted in BJD and are fitted with the same quadratic ephemeris as above. The results of The Monte Carlo are: \\
$T_{rot} = 2444782.8967 \pm 0.0016$ BJD \\
$P_{rot} = 1254.48379 \pm 0.00092$ s \\
$b_{rot} = -1.002.10^{-12} \pm 0.021.10^{-12}$ day \\
The resulting O-C diagram is shown in Figure 3.
\\ \\
\epsfysize=8.5cm
\centerline{\epsffile{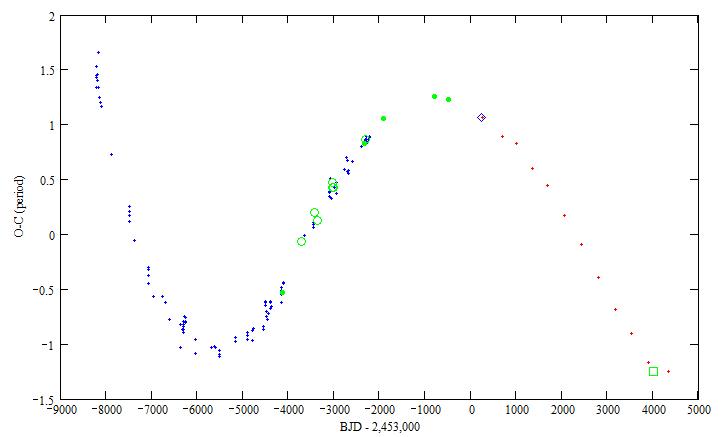}}
\textit{Figure 3: Blue dots: the data from P98, Green dots: the W03 observations, Green circles: Kruszewski \& Semeniuk, 1998, Blue diamond: Andronov et al, 2005, Green square: Kennedy et al, 2016, Red dots: this work.}
\\ \\
The period of rotation of the white dwarf is decreasing at a rate $P_{rot}'= 2b_{rot}/P_{rot} = -1.380.10^{-10} \pm 0.029.10^{-10}$, over a time scale $P_{rot}/2P_{rot}' = -144$ kyr (about the same as for AO Psc, see Bonnardeau, 2015). The O-C diagram may suggest an oscillation with a time scale of about 25 yr; this is not due to a third body as the residuals for the orbital ephemeris show no modulation.
\\ \\
\textbf{}
\\
\textbf{Amplitude variations} \\

The variations of the amplitudes do not show any obvious trend or correlation, except for $A_{rot}$ which stayed nearly constant at -0.2 mag from 2004 to 2011, then reached -0.35 mag in 2015, as shown in Figure 4.
\\
\epsfysize=11.5cm
\centerline{\epsffile{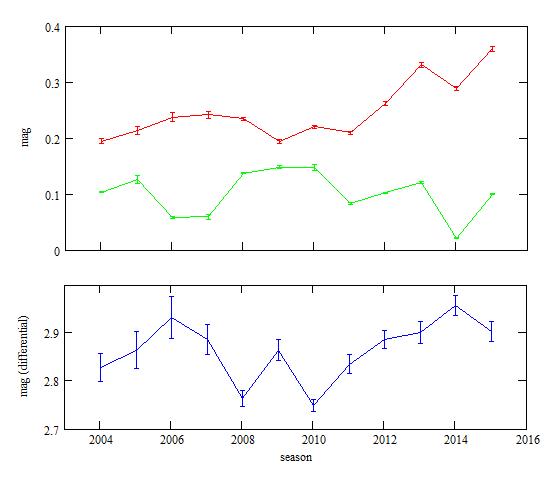}}
\textit{Figure 4: Red: $-A_{rot}$, Green: $A_{orb}$, Blue: $A_{0}$.}  
\\ \\
However, a preliminary result for the season 2016 is that FO Aqr has fainted by about 1.5 mag; see also Littlefield et al, 2016. 
\\ \\ \\
\textit{Acknowledgment:} The use of the on-line tool of the University of Ohio to convert HJD to BJD, at 
http://astroutils.astronomy.ohio-state.edu/time/hjd2bjd.html, is acknowledged. 
\\ 
\textbf{}
\\
\textbf{References} \\ 

Andronov I.L., Ostrova N.I. and Burwitz V., 2005, \textit{ASP Conf. Series} \textbf{335} 229.

Bonnardeau M., 2015, \textit{IBVS} 6146. 

Kennedy M.R. et al, 2016, \textit{MNRAS} \textbf{459} 3622.

Kruszewski A. and Semeniuk I., 1998, \textit{Acta Astronomica} \textbf{48} 757. 

Littlefield C. et al, 2016, \textit{ATEL} 9216 and 9225. 

Patterson J. et al, 1998, \textit{PASP} \textbf{110} 415.

Williams G., 2003, \textit{PASP} \textbf{115} 618.

\end{document}